\documentclass[%
prereprint,
showpacs,preprintnumbers,
 amsmath,amssymb,
 aps,
 prl,
]{revtex4-1}

\usepackage{graphicx}
\usepackage{dcolumn}
\usepackage{bm}

\begin{document}


\title{Quasirelativistic Potential Energy Curves of NaRb for Direct Spectra Interpretation}


\author{M. Wiatr}
\affiliation{Department of Theoretical Physics and Quantum Information,\\ Faculty of Applied Physics and Mathematics, Gdansk University of Technology,\\ ul. Gabriela Narutowicza 11/12, 80-233 Gdansk, Poland}
\author{P. Jasik}
\affiliation{Department of Theoretical Physics and Quantum Information,\\ Faculty of Applied Physics and Mathematics, Gdansk University of Technology,\\ ul. Gabriela Narutowicza 11/12, 80-233 Gdansk, Poland}
\author{T. Kilich}
\affiliation{Department of Theoretical Physics and Quantum Information,\\ Faculty of Applied Physics and Mathematics, Gdansk University of Technology,\\ ul. Gabriela Narutowicza 11/12, 80-233 Gdansk, Poland}
\author{H. Stoll}
\affiliation{Institute for Theoretical Chemistry,\\ University of Stuttgart,\\  Pfaffenwaldring 55, D-70569 Stuttgart, Germany}
\author{J.E. Sienkiewicz}
\email{jes@mif.pg.gda.pl}
\affiliation{Department of Theoretical Physics and Quantum Information,\\ Faculty of Applied Physics and Mathematics, Gdansk University of Technology,\\ ul. Gabriela Narutowicza 11/12, 80-233 Gdansk, Poland}

\date{\today}

\begin{abstract}
We report the quasirelativistic potential energy curves including spin-orbit and scalar-relativistic effects of the NaRb molecule. The calculated curves of the 0$^+$, 0$^-$, 1, and 2 molecular states correlate for large internuclear separation with the five lowest atomic energies covering the first P doublet states of the rubidium and sodium atoms. Several new features of potential curves are found. Some disagreements with the other theoretical data are found and discussed. We advocate that our results may be used in the direct way, skipping the unperturbation of the spin-orbit interaction, for assigning the molecular spectra.
\end{abstract}

\maketitle


{\it Introduction.--} There has been a long interest in the experimental and theoretical studies of alkali heteronuclear dimer molecules, since they are relatively simple electronic systems which molecular spectra coming from low-lying states can be precisely resolved. Moreover, the permanent electric dipole moment makes them one of the best candidates for slowing and trapping using inhomogeneous electric field, hence enabling to study quantum phenomena as Bose condensation and Fermi superfluidity.

In the past twenty years, several spectroscopic studies on the NaRb dimer were reported. Experiments used different techniques like the polarization labeling spectroscopy \cite{2006Pashov}, the Fourier transform spectroscopy of laser induced fluorescence supported by the deperturbation treatment of the A$^1\Sigma^+$\slash b$^3\Pi$ complex \cite{2002Tamanis, 2007Docenko}, the Doppler-free optical-optical double resonance polarization spectroscopy \cite{1996Kasahara}, and very recently the stimulated Raman adiabatic passage \cite{2016Zho, 2016Guo}.

Theoretical studies of the electronic structure of NaRb include the application of the many-body multipartitioning perturbation theory \cite{2001Zaitsevskii}, the configuration interaction by perturbation of the multiconfiguration wave function method \cite{2000Korek, 2012Dardouri, 2014Chaieb}, and our recent calculations based on the multiconfigurational complete active space self-consistent field and multireference configuration interaction methods \cite{2015Wiatr}. Only one approach (Korek et al. \cite{2009Korek}), using the CIPSI codes, dealt with the relativistic spin-orbit effect taken into account by use of the semiempirical spin-orbit pseudopotential added to the electrostatic Hamiltonian. However, additional theoretical results obtained by independent computational methods including relativistic effects are required to address the need for accurate differences between electronic and rovibrational levels which are very essential in laser tunings and resolving measured spectra.

The aim of this work is to provide quasirelativistic potential energy curves including the spin-orbit and scalar-relativistic effects for low-lying states of NaRb. The atomic asymptotes and molecular parameters will be compared with other available theoretical results. To assess very demanding and time-consuming tasks, we would like to draw the reader's attention to the direct assignment of experimental spectra with the help of potential energy curves given for the Hund's case (c). In our opinion, this approach is much simpler than the use of potential energy curves calculated without the spin-orbit couplings. Particularly, in the case of where a strong singlet/triplet mixture requires the usage of computationally demanding deperturbation methods with several model parameters. Our proposal will be supported by comparison with the adequate example of a recent experiment on the high resolution Fourier transform spectra of NaRb molecules.

{\it Method.--} In our computational approach, the alkali dimer is considered as an effective two-electron system. Each atom is replaced by one valence electron and the core consisting of a point nucleus and remaining electrons from the closed atomic subshells. Since the theoretical approach, without the spin-orbit coupling, has been already presented in the earlier paper \cite{2015Wiatr}, here we give only salient details concerning the spin-orbit pseudopotential and atomic basis sets. The calculations are based on the multireference configuration interaction (MRCI) method with the atomic effective core potentials and the core polarization potentials, which enable to treat explicitly only two valence electrons of the NaRb dimer. A rich atomic orbital basis allows to obtain reliable results of potential energy curves of chosen molecular states. Calculations of potential energy curves are performed by means of the MOLPRO program package \cite{2012Werner}, while all spectroscopic parameters are obtained by the Level16 program \cite{2017LeRoy}. The core electrons of sodium atom are represented by the ECP10SDF pseudopotential \cite{1982Fuentealba}. In the case of s and p functions, we use the basis set for sodium which comes with the ECP10SDF pseudopotential. In turn, for the d and f functions we use the cc-pVQZ basis set \cite{2011Prascher}. Additionally, these basis sets are augmented by thirteen s functions, six p functions, seven d functions, and two f functions. For rubidium atom, the core electrons are represented by the scalar-relativistic effective core pseudopotential  \cite{1986Silberbach}. Here, the set of s and p functions coming with the ECP36SDF pseudopotential \cite{1982Szentpaly, 1983Fuentealba} is expanded by the set of d and f functions coming with the effective core potential ECP28MDF \cite{2005Lim} and augmented by thirteen s functions, six p functions, seven d functions, and two f functions. All exponents of the augmented Gaussian functions can be found in our earlier paper \cite{2015Wiatr}.

The scalar relativistic effects are described by the energy-consistent effective core pseudopotentials. The spin-orbit operator is included by means of the spin-orbit pseudopotentials which are taken in the following l-dependent form \cite{2001Czuchaj}
\begin{equation}
V_{SO, \lambda} = \sum_{i, l}\frac{2\Delta V_{i,\lambda l}}{2l+1} P_{\lambda l}{\bf r}_i\cdot {\bf l}_i P_{\lambda l},
\end{equation}
where $P_{\lambda l} = \sum_{m=-l}^l |\lambda lm><\lambda lm|$ is the projection operator onto Hilbert subspace of angular symmetry $l$ with respect to core $\lambda$, i indicates the i-th valence electron of molecule, and l is the orbital quantum number. Here, the index $\lambda$, assigned Na or Rb, goes over the atomic cores of sodium and rubidium atoms. The difference $\Delta V_{i,\lambda l}$ of the radial parts of the two component pseudopotentials $V_{\lambda l, l+1/2}$ and $V_{\lambda l, l-1/2}$ is given in terms of Gaussians functions
\begin{equation}
\Delta V_{i,\lambda l}= \sum_k A_{\lambda, lk} exp(-a_{\lambda, lk} r^2_{\lambda i}).
\end{equation}
The spin-orbit pseudopotential parameters $A_{\lambda, lk}$ and $a_{\lambda, lk}$ for sodium and rubidium are given by Soorkia et al. \cite{2007Soorkia} and Silberbach et. al. \cite{1986Silberbach}, respectively. In the latter case, we modified two parameters, namely A$_{Rb, 11}$ and  A$_{Rb, 12}$ in order to have better adjustments with the experimental values of the spin-orbit splitting (TAB. \ref{tab:so_rubid_popr}). The calculated spin-orbit matrix elements are added to the Hamiltonian matrix at the MRCI level.

\begin{table}[!h]
\caption{\label{tab:so_rubid_popr}The spin-orbit pseudopotential parameters (in atomic units) for the Rb atom from Silberbach et. al \cite{1986Silberbach} with two modified A$_{Rb, 11}$ and A$_{Rb, 12}$ parameters.}
\begin{ruledtabular}
\begin{tabular}{ccccc}
$lk$&11&12&21&22\\
\hline
$A_{Rb, lk}$&-2.9502&2.8551&1.9574&-1.8972\\
$a_{Rb, lk}$&0.31777&0.30313&0.37996&0.37277\\
\end{tabular}
\end{ruledtabular}
\end{table}

The potential energy curves are computed using the multiconfigurational self-consistent field/complete active space self-consistent field (MCSCF/CASSCF) method to generate the orbitals for the subsequent CI calculations. The corresponding active space involves the molecular counterparts of the 3s, 3p, and 3d valence orbitals of sodium as well as 5s, 5p, and 4d valence orbitals of rubidium. All 18 spd orbitals are included in our calculations. 

\begin{figure*}
\includegraphics[scale=0.5]{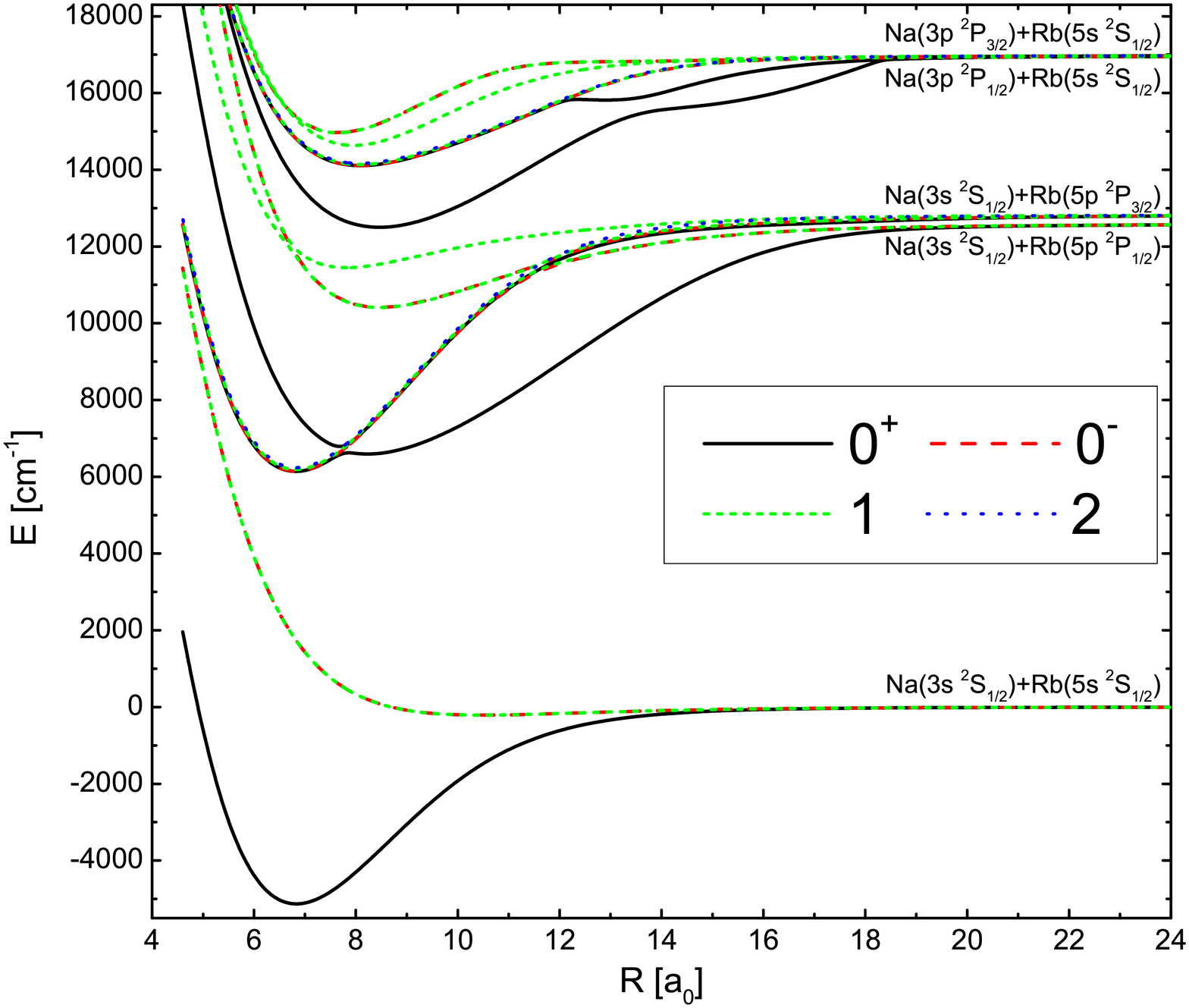}
\caption{\label{fig:all_states}(Color online) The NaRb lowest potential energy curves including spin-orbit interaction.}
\end{figure*}

{\it Results and discussion.--}The molecular calculations were performed for internuclear distance R in the range from 4.6 to 86 a$_0$ with different step sizes. Calculated potential energy curves correlate for infinite R to the five following combinations of atomic states: Na(3s~$^2$S$_{1/2}$) + Rb(5s~$^2$S$_{1/2}$), Na(3s~$^2$S$_{1/2})$ + Rb(5p~$^2$P$_{1/2})$, Na(3s~$^2$S$_{1/2})$ + Rb(5p~$^2$P$_{3/2})$, Na(3p~$^2$P$_{1/2})$ + Rb(5s~$^2$S$_{1/2})$, and Na(3p~$^2$P$_{3/2})$ + Rb(5s~$^2$S$_{1/2})$. We routinely check the quality of our basis sets by performing the CI calculations for the ground and excited states of both atoms. In TAB. \ref{tab:asymptotes}, the present asymptotic energies show very good and consistent agreement with atomic levels given by Sansonetti \cite{2006Sansonetti, 2008Sansonetti}, while Korek and Fawwaz \cite{2009Korek} achieved very good agreement only for the first two excited levels of Rb.
\begin{table*}
\caption{\label{tab:asymptotes}The comparison of asymptotic energies with other theoretical and experimental results. $\Delta_{S-p}$ and $\Delta_{S-K}$ stand for differences between experimental values provided by Sansonetti \cite{2006Sansonetti,2008Sansonetti} and present as well as Korek and Fawwaz \cite{2009Korek} results, respectively. Energies are shown in cm$^{-1}$ units. The capital letter T refers to theoretical results and E denotes experimental data.}
\begin{ruledtabular}
\begin{tabular}{cccccc}
Asymptotes&\mbox{present T}&\mbox{Sansonetti E}&\mbox{Korek and Fawwaz T}&\mbox{$\Delta_{S-p}$}&\mbox{$\Delta_{S-K}$}\\
\hline
$Na(3s^{2}S_{1/2})+Rb(5p^{2}P_{1/2})$&12579.55&12578.95&12578.32&-0.60&0.63\\
$Na(3s^{2}S_{1/2})+Rb(5p^{2}P_{3/2})$&12818.89&12816.55&12815.99&-2.34&0.56\\
$Na(3p^{2}P_{1/2})+Rb(5s^{2}S_{1/2})$&16953.82&16956.17&16967.48&2.35&-11.31\\
$Na(3p^{2}P_{3/2})+Rb(5s^{2}S_{1/2})$&16973.64&16973.37&17004.41&-0.27&-31.04\\
\end{tabular}
\end{ruledtabular}
\end{table*}	
Our results of potential energy curves are displayed in FIG. \ref{fig:all_states}. The potential energy curve of the (1)0$^+$ ground state has the shape of the Morse potential. An important feature for the photoassociation process is the avoided crossing between the potential energy curves of the (2)0$^+$ and (3)0$^+$ states which takes place around 7.8 a$_0$. It results in the double well of the lower curve and subsequently in the irregular spacings of vibrational levels. The upper curve assumes the shape of the Morse potential. Another interesting feature is the interaction between potential curves of the (4)0$^+$ and (5)0$^+$ states which respectively correlate with P$_{1/2}$ and P$_{3/2}$ of sodium doublet. This interaction leads to avoided crossings at R = 13.8 and 18.4 a$_0$. The wavy shape of the (5)0$^+$ potential curve comes from several avoiding crossings with the higher (6)0$^+$ potential curve (not shown in FIG. \ref{fig:all_states}) correlating with Na(3s $^2$S$_{1/2})$ + Rb(4d $^2$D$_{5/2})$ atomic asymptote. At R = 11.8 a$_0$ there are avoided crossings between potential energy curves of the (2)0$^-$ and (3)0$^-$ states as well as between the (2)1 and (3)1 states. At smaller distance R = 6.8 a$_0$ the potential curve of the (3)1 state avoids crossing with the potential curve of the (4)1 state. 

\begin{figure*}
\includegraphics[scale=0.5]{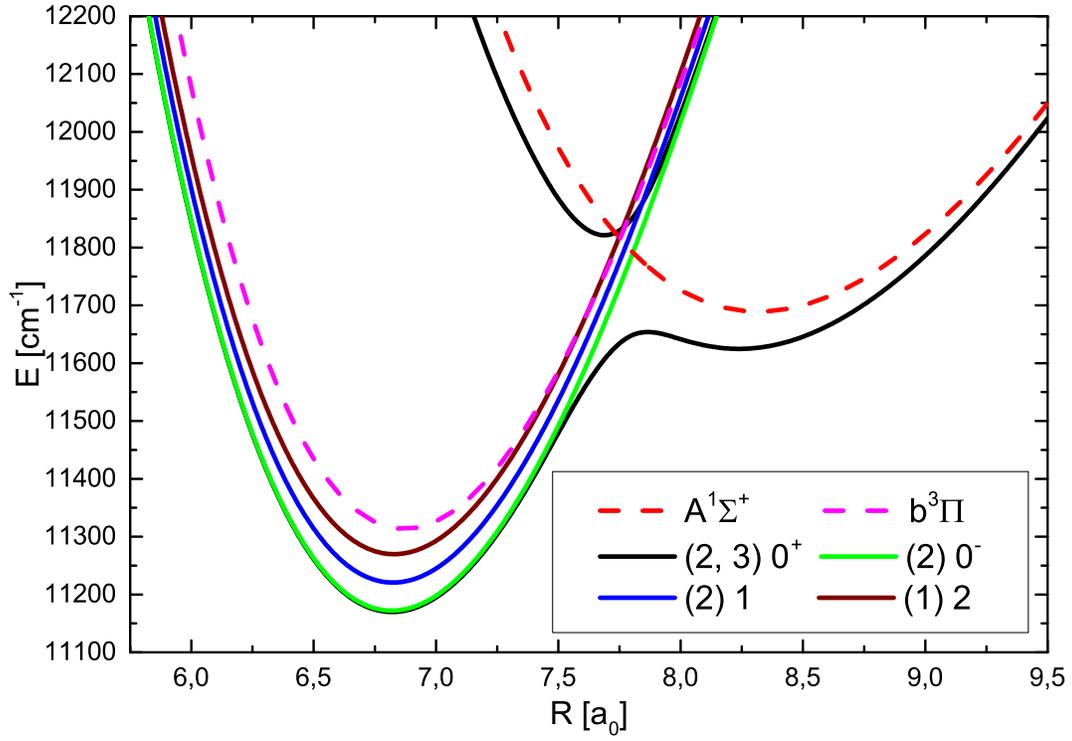}
\caption{\label{fig:comp_qrel_Docenko}(Color online) The comparison of the excited quasirelativistic potentials for the $(2, 3) 0^+$, $(2) 0^-$, $(2) 1$, and $(1) 2$ states of the NaRb molecule with deperturbed $A^{1}\Sigma^{+}$ and $b^{3}\Pi$ states derived from experimental data by Docenko et al. \citep{2007Docenko}.}
\end{figure*}

The presently calculated spectroscopic parameters along with ones from theoretical work of Korek and Fawwaz \cite{2009Korek} as well as results derived from the experiment using deperturbation techniques given by Docenko et al. \cite{2007Docenko} are shown in TAB. \ref{tab:parameters}.
\begin{table*}
\caption{\label{tab:parameters} The spectroscopic parameters of the bond length $R_e$ ($a_0$), dissociation energy $D_e$, electronic term energy $T_{e}$, dissociation limit referred to the minimum of the ground state $T_{dis}$, vibrational constant $\omega_{e}$, and rotational constant $B_e$ ($cm^{-1}$) for the excited states of the NaRb molecule.}
\centering
\begin{ruledtabular}
\begin{tabular}{@{}*{10}{r}}
 \multicolumn{1}{c}{State} & \multicolumn{1}{c}{$R_e$} & \multicolumn{1}{c}{$D_e$} & \multicolumn{1}{c}{$T_{e}$} & \multicolumn{1}{c}{$T_{dis}$} & \multicolumn{1}{c}{$\omega_{e}$} & \multicolumn{1}{c}{$B_e$} & \multicolumn{1}{c}{Rubidium asymptote} & \multicolumn{1}{c}{} \\
\hline
 &  &  &  &  &  &  &  &  & \\
       \multicolumn{1}{c}{$A^{1}\Sigma^{+}$} & 8.315 & 6080 & 11688 & 17768 &       &          & \multicolumn{1}{c}{$5p ^{2}P$}       & \multicolumn{1}{l}{Docenko et al. \cite{2007Docenko}} \\
\multicolumn{1}{c}{$(2) 0^{+}$ (outer well)} & 8.220 & 5989 & 11731 & 17721 & 65.55 & 0.050821 & \multicolumn{1}{c}{$5p ^{2}P_{1/2}$} & \multicolumn{1}{l}{present} \\
			                                 & 8.200 &  & 11765 &       &       &          &                                      & \multicolumn{1}{l}{Korek et al. \cite{2009Korek}} \\
\hline
 &  &  &  &  &  &  &  &  & \\
\multicolumn{1}{c}{$b^{3}\Pi_{0}$} & 6.869 & 6378 & 11361 & 17689 &       &          & \multicolumn{1}{c}{$5p ^{2}P_{1/2} + \xi_{Rb}^{so}$} & \multicolumn{1}{l}{Docenko et al. \cite{2007Docenko}} \\
\multicolumn{1}{c}{$(2) 0^{+}$ (inner well)} & 6.816 & 6451 & 11270 & 17721 & 104.18 & 0.071522 & \multicolumn{1}{c}{$5p ^{2}P_{1/2}$} & \multicolumn{1}{l}{present} \\   									             & 6.767 & 6578\footnote{According to us this value describes the dissociation energy D$_e$ of the inner well of the (2)$0^+$ state, not as reported in \cite{2009Korek} D$_e$ of the outer well.} & 11259 &       & 105.70 & 0.072590 &                                      & \multicolumn{1}{l}{Korek et al. \cite{2009Korek}} \\
	         \multicolumn{1}{c}{$(2) 0^{-}$} & 6.814 & 6449 & 11272 & 17721 & 104.45 & 0.071556 & \multicolumn{1}{c}{$5p ^{2}P_{1/2}$} & \multicolumn{1}{l}{present} \\
		                                     & 6.765 & 6577 & 11263 &       & 105.90 & 0.072610 &                                      & \multicolumn{1}{l}{Korek et al. \cite{2009Korek}} \\
			     \multicolumn{1}{c}{$(2) 1$} & 6.818 & 6400 & 11320 & 17721 & 104.42 & 0.071455 & \multicolumn{1}{c}{$5p ^{2}P_{1/2}$} & \multicolumn{1}{l}{present} \\
			                                 & 6.761 & 6530 & 11310 &       & 106.00 & 0.072690 &                                      & \multicolumn{1}{l}{Korek et al. \cite{2009Korek}} \\
			     \multicolumn{1}{c}{$(1) 2$} & 6.824 & 6591 & 11369 & 17960 & 104.40 & 0.071355 & \multicolumn{1}{c}{$5p ^{2}P_{3/2}$} & \multicolumn{1}{l}{present} \\
			                                 & 6.758 & 6720 & 11358 &       & 106.20 & 0.072770 &                                      & \multicolumn{1}{l}{Korek et al. \cite{2009Korek}} \\
\hline
 &  &  &  &  &  &  &  &  & \\
			 \multicolumn{1}{c}{$(3) 0^{+}$} & 7.706 & 6061 & 11899 & 17960 & 175.10 & 0.056488 &  \multicolumn{1}{c}{$5p ^{2}P_{3/2}$} & \multicolumn{1}{l}{present} \\
            \end{tabular}
		\end{ruledtabular}
\end{table*}
The most important differences between these three sets of parameters are connected with the double minimum of the (2)$0^+$ state rising due to the avoided crossing with the (3)$0^+$ state. The outer well of this potential in the Hund's case (c) corresponds to $A^{1}\Sigma^{+}$ in the Hund's case (a), while the inner well is an equivalent to $b^{3}\Pi$. The relative positions of present quasirelativistic potential energy curves and potentials derived from experimental results by the deperturbation procedure \cite{2007Docenko} are shown in FIG. \ref{fig:comp_qrel_Docenko}. It should be noticed that in the Hund's case (a) considered potential energy curves of the NaRb molecule correlate in the asymptotic region to the center of the multiplet Rb($5p ^{2}P$), while in the Hund's case (c) proper atomic limits are Rb($5p ^{2}P_{1/2}$) and Rb($5p ^{2}P_{3/2}$). Only limited comparison between chosen spectroscopic parameters coming from these two representations can be performed. For the equilibrium length R$_e$, differences between experimental value of Docenko \cite{2007Docenko} and present as well as Korek and Fawwaz \cite{2009Korek} results amount to 0.095 and 0.115 a$_0$, respectively. Present dissociation energy D$_e$ of the outer well of the (2)0$^+$ state equals to 5989 cm$^{-1}$, whereas Korek and Fawwaz \cite{2009Korek} do not determine this parameter. The relative depth of the outer well of the (2)0$^+$ state, taken as a difference between the top of the barrier and the minimum value, equals to 53 cm$^{-1}$. The same parameter given by Korek and Fawwaz \cite{2009Korek} amounts only to 29 cm$^{-1}$. In our case, the well depth is sufficient to localize rovibrational levels and allows to determine vibrational constant $\omega_e$ = 65.55 cm$^{-1}$ and rotational constant B$_e$ = 0.050821 cm$^{-1}$.

The case of the inner well of the (2)$0^+$ state is more complicated, because of the existence of additional potential energy curves of the (2)0$^-$, (2)1, and (1)2 states. The splitting between these potentials around the minimum of the inner well (FIG. \ref{fig:comp_qrel_Docenko}) almost exactly obeys the formula for the electronic energy of a multiplet term $T_e = T_0 + A^{SO}\Lambda\Sigma$, where $T_0$ is the term value when the spin is neglected, $\Lambda$ is the component of the electronic orbital angular momentum along the internuclear axis, and $A^{SO}$ is the spin-orbit constant for a given multiplet term \cite{1989Herzberg}. The values of $A^{SO}$ in the considered system are 50.6 and 49.1 cm$^{-1}$ for $E^{(2)1}(R_e) - E^{(2)0^+}(R_e)$ and $E^{(1)2}(R_e) - E^{(2)1}(R_e)$, respectively. These parameters are in good agreement with the results given by Korek and Fawwaz \cite{2009Korek}, which are equal to 50.9 and 48.2 cm$^{-1}$, respectively. We are able to compare present spectroscopic parameters of the inner wells of the (2)$0^+$ and (2)0$^-$ states with the component $\Omega = 0$ of the $b^{3}\Pi$ state reported by Docenko at al. \cite{2007Docenko}. For instance, differences between the experimental value of the equilibrium length R$_e$ and ours is around 0.05 a$_0$ and is twice as small as the difference between the experimental value and the result given by Korek et al. \cite{2009Korek}. It is visible from TAB. \ref{tab:parameters}, that the present dissociation energies D$_e$ calculated for the inner wells of the (2)$0^+$ and (2)$0^-$ states are respectively 73 and 71 cm$^{-1}$ larger in the comparison with the value reported by Docenko et al. \cite{2007Docenko}. The same well depths given by Korek and Fawwaz \cite{2009Korek} are respectively deeper by 200 and 199 cm$^{-1}$, than the experimental value. It can be noticed that in the asymptotic region the $b^{3}\Pi_0$ state reported by Docenko et al.   correlates with Rb($5p ^{2}P_{1/2}$) + $\xi_{Rb}^{so}$ instead with Rb($5p ^{2}P_{1/2}$). Our rubidium spin-orbit splitting $\xi_{Rb}^{so} = [E_{5p ^{2}P_{3/2}} - E_{5p ^{2}P_{1/2}}]/3$ is equal to 79.78 cm$^{-1}$ and agrees very well with experimental data provided by Sansonetti (79.20 cm$^{-1}$) \cite{2006Sansonetti,2008Sansonetti} and Docenko et al. (79.22 cm$^{-1}$) \cite{2007Docenko}. 

\begin{figure*}
\includegraphics[scale=0.45]{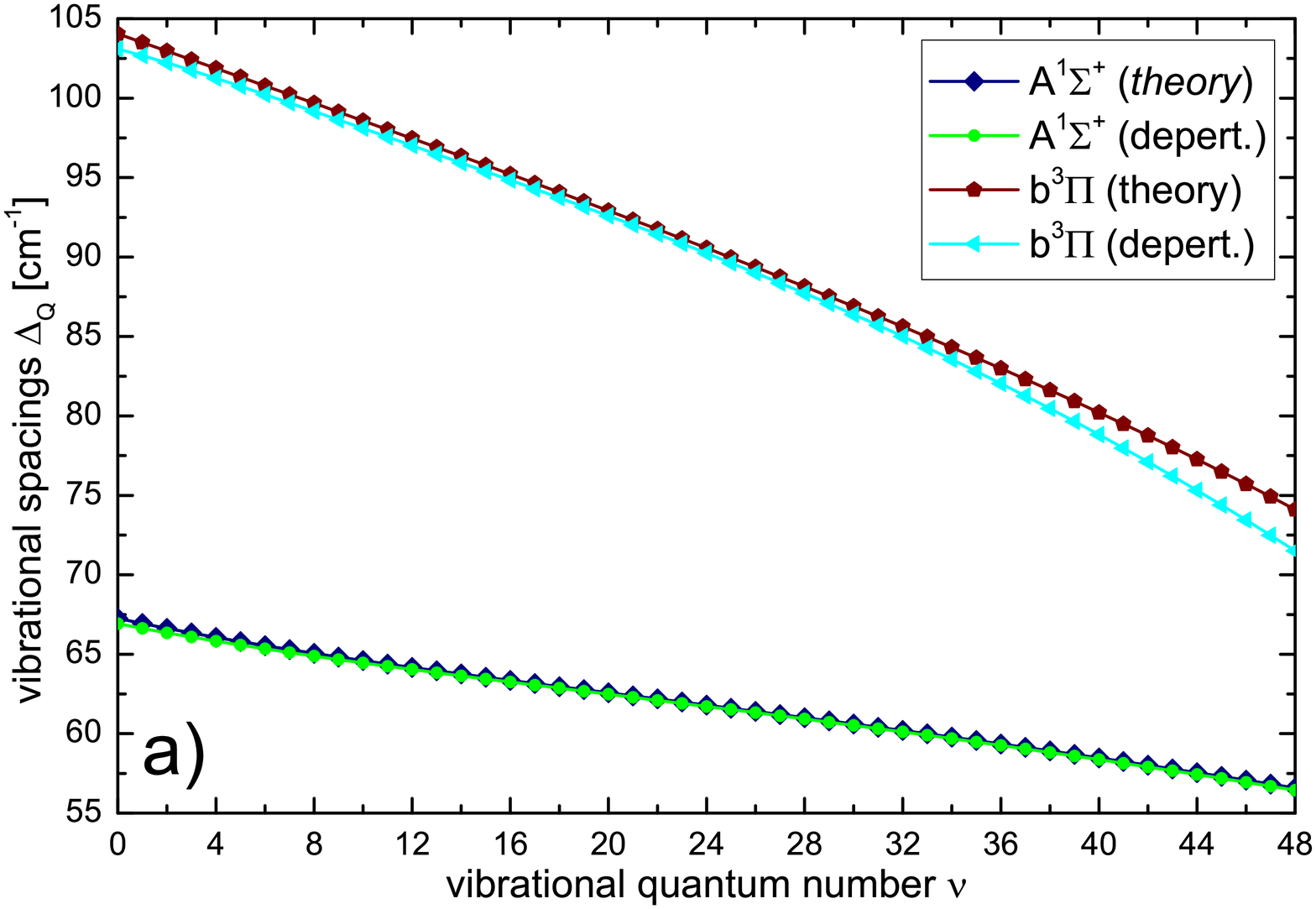} \\
\includegraphics[scale=0.45]{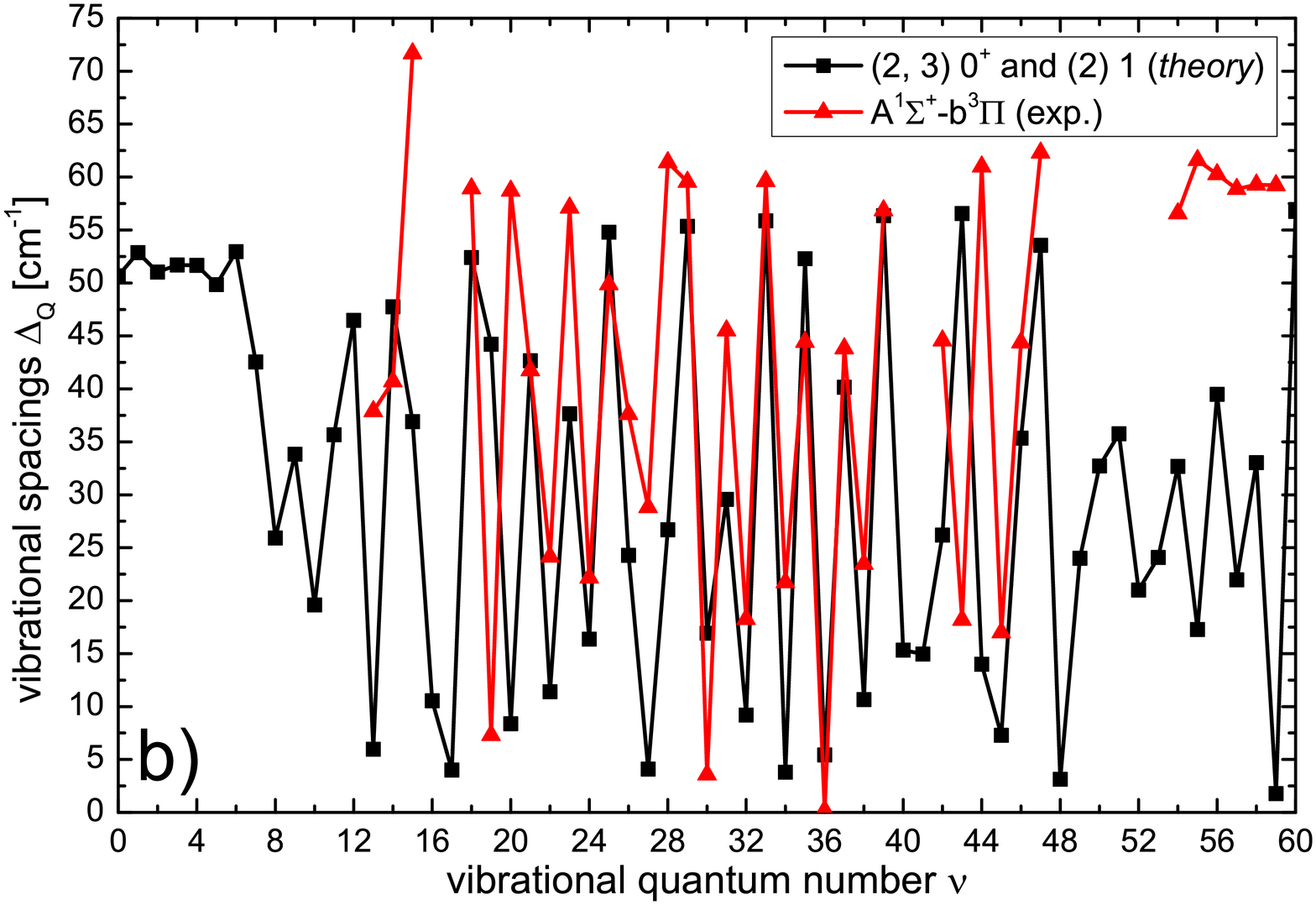}
\caption{\label{fig:spacings}(Color online) Vibrational spacings of the $^{23}$Na$^{85}$Rb molecule. (a) Comparison of the regular vibrational spacings $\Delta_Q = \nu(v", J"=0) - \nu(v" +1, J"=0)$ of the A$^1\Sigma^+$ and b$^3\Pi$ states between theoretical \cite{2015Wiatr} and deperturbed \cite{2007Docenko} values. (b) Irregular vibrational spacings $\Delta_Q = \nu(v", J"=25) - \nu(v" +1, J"=25)$ of the presently calculated (2)0$^+$, (3)0$^+$, and (2)1 states compared with the experimental spacings of the Q branch for D$^1\Pi \rightarrow$ A$^1\Sigma^+$\slash b$^3\Pi$
progression taken from supplementary material \cite{2007Docenko}.}
\end{figure*}

Quasirelativistic results of the potential energy curves may be helpful to resolve in detail the NaRb spectra and to find efficient schemes for the ultracold molecule formation in the deeply bound ground state. One of the difficult cases is the resolution of laser induced fluorescence spectra of transitions from higher excited states like D$^1\Pi$ to the A$^1\Sigma^+$\slash b$^3\Pi$ complex, reported by Docenko at al. \cite{2007Docenko}. The spin-orbit coupling between singlet and triplet states of this A-b complex is responsible for strong irregularities in the spacings of vibrational and rotational levels. In FIG. \ref{fig:spacings}(a), we compare the vibrational spacings calculated from potential energy curves reported in our previous paper \cite{2015Wiatr} with the one obtained experimentally \cite{2007Docenko}. As expected, there is a very good agreement between spacings coming from the A$^1 \Sigma ^+$ state and these calculated from the potentials derived from experimental spectra using the spin-orbit deperturbation approach. The same may be said about analogical comparison concerning the b$^3\Pi$ state, except the region with higher values of the vibrational quantum number $v$ where some slowly growing disagreement is visible.

A quite different situation occurs when it comes to the comparison between experimental and present theoretical spacings resulting from our quasirelativistic potential energy curves (FIG. \ref{fig:spacings}(b)). The enumeration of the rovibrational levels of the A$^1\Sigma^+$\slash b$^3\Pi$ complex reported by Docenko et al. \cite{2007Docenko} starts from $v=0$ for the A$^1\Sigma^+$ state. Moreover, all rovibrational levels are presented as a percentage share of A$^1\Sigma^+$, b$^3\Pi_0$, b$^3\Pi_1$, and b$^3\Pi_2$ components. In our approach, electronic states are represented by separate quasirelativistic potential energy curves (FIG. \ref{fig:comp_qrel_Docenko}), so we are able to precisely assign the origin of each rotational and vibrational levels. The beginning of enumeration of the rovibrational levels is natural and it is connected with the inner well of the (2)$0^+$ state (FIG. \ref{fig:comp_qrel_Docenko}). As an example, we take spacings between experimentally given rovibrational levels assigned to A$^1\Sigma^+$ and b$^3\Pi$ states with $J"=25$ and compare them with the most appropriate vibrational spacings of our (2)0$^+$, (3)0$^+$, and (2)1 states. The (1)2 state is omitted because the percentage shared of the b$^3\Pi_2$ component in the spectrum is negligible. The comparison displayed in FIG. \ref{fig:spacings}(b) shows significant pattern resemblance. In this figure, we consequently use our numbering of $v$ and we omit unassigned parts of the experimental spectrum. Knowing how many rovibrational levels of (2)0$^+$, (3)0$^+$, and (2)1 states are lying below A$^1\Sigma^+(v"=2, J"=25)$ it is possible to find the matching point of the two representations. At this stage, it is rather difficult to exactly match both spacings since even small assignment ambiguities on the experimental side and small inaccuracies on the theoretical side give progressive discrepancies. Nevertheless, we noticed that our quasirelativistic potentials may be more convenient to use than nonrelativistic ones, if taken as the starting potentials for inverted perturbation approach. Moreover, in this case there will be no need to perform spin-orbit deperturbation. The spectra of heavier alkali molecules may be directly interpreted using quasirelativistic potential energy curves.

{\it Conclusion.--} This paper presents the results of quasirelativistic calculations of adiabatic Born-Oppenheimer potential energy curves of the NaRb molecule. In our computational approach only two valence electrons are taken explicitly to build determinants for the multiconfiguration wave function. Atomic cores are described by local pseudopotentials and core polarization potentials. Taken into account relativistic effects include spin-orbit interaction and scalar-relativistic effect. Calculated potential energy curves very accurately correlate with (1) $^2S_{1/2}$ ground state of Na and Rb atoms, (2) ground state $^2S_{1/2}$ of the Na atom and the first excited doublet states $^2P_{1/2, 3/2}$ of the Rb atom, and (3) the first excited doublet states $^2P_{1/2, 3/2}$ of the Na atom and ground state $^2S_{1/2}$ of the Rb atom. 

Several new features are described, among them avoided crossings between potential energy curves of the states with $0^+$ symmetry. Very good agreement is found between the vibrational spacings calculated and derived from experiment deperturbed A$^1\Sigma^+$ and b$^3\Pi$ states. Example irregularities in the experimental rovibrational spacings of the strongly perturbed D$^1\Pi$ $\rightarrow$ A$^1\Sigma^+$\slash b$^3\Pi$ progression are explained by direct comparison with the spacings of joined (2)$0^+$, (3)$0^+$, and (2)$1$ states. In conclusion, we suggest the direct use of quasirelativistic potentials for the resolution of an assignment of experimental spectra of diatomic molecules.

{\it Acknowledgements.--}We thank Olivier Dulieu, Wlodzimierz Jastrzebski, Pawel Kowalczyk, and Jacek Szczepkowski for useful discussion. We acknowledge partial support by the COST Action CM1204 of the European Community and the Polish Ministry of Science and Higher Education. Calculations have been carried out using resources provided by Academic Computer Centre in Gdansk and Wroclaw Centre for Networking and Supercomputing. Proofreading by Ashley Eriksen is gracefully acknowledged.

\newpage

\end{document}